\documentclass[sigconf]{acmart}

\usepackage{graphicx}
\usepackage{multirow}
\usepackage{lipsum}
\usepackage{subcaption}
\usepackage[para]{footmisc}
\usepackage[normalem]{ulem}
\usepackage{booktabs}
\usepackage{amsmath}
\usepackage{tcolorbox}
\usepackage[T1]{fontenc}
\usepackage[utf8]{inputenc}
\usepackage{color}
\usepackage{array}
\usepackage{algorithm}
\usepackage{algpseudocode}
\usepackage{marginnote}
\usepackage[show]{chato-notes}

\newcommand{\cm}[1]{\textcolor{black}{#1}}
\newcommand{\maria}[1]{\textcolor{black}{#1}}
\newcommand{\mv}[1]{\textcolor{black}{#1}}

\newcommand{\pageenlarge}[1]{\enlargethispage{#1\baselineskip}}

\DeclareMathOperator*{\argmax}{arg\,max}

\setcopyright{acmlicensed}
\copyrightyear{2025}
\acmYear{2025}
\acmDOI{10.1145/3705328.3748149}
\acmConference[RecSys '25]{19th ACM Conference on Recommender Systems}{September 22--26,
  2025}{Prague, Czech Republic}
\acmBooktitle{19th ACM Conference on Recommender Systems (RecSys ’25),
  September 22--26, 2025, Prague, Czech Republic}
\acmISBN{979-8-4007-1364-4/2025/09}

\begin{document}

\title[Fashion-AlterEval: A Dataset for Evaluating CRS with Alternatives]{Fashion-AlterEval: A Dataset for Improved Evaluation of Conversational Recommendation Systems with Alternative Relevant Items}


\author{Maria Vlachou}
\affiliation{%
  \institution{University of Glasgow}
  \city{Glasgow}
  \country{UK}}
\email{m.vlachou.1@research.gla.ac.uk}



\renewcommand{\shortauthors}{Vlachou}

\begin{abstract}
 In Conversational Recommendation Systems (CRS), a user provides feedback on recommended items at each turn, leading the CRS towards improved recommendations. Due to the need for a large amount of data, a user simulator is employed for both training and evaluation. Such user simulators critique the current retrieved item based on knowledge of a single {\em target} item. However, system evaluation in offline settings with simulators is limited by the focus on a single target item and their unlimited patience over a large number of turns. To overcome these limitations of existing simulators, we propose \textit{Fashion-AlterEval}, a new dataset that contains human judgments for a selection of alternative items by adding new annotations in common fashion CRS datasets. Consequently, we propose two novel meta-user simulators that use the collected judgments and allow simulated users not only to express their preferences about alternative items to their original target, but also to change their mind and level of patience. In our experiments using the Shoes and Fashion IQ as the original datasets and three CRS models, we find that using the knowledge of alternatives by the simulator can have a considerable impact on the evaluation of existing CRS models, specifically that the existing single-target evaluation underestimates their effectiveness, and when simulated users are allowed to instead consider alternative relevant items, the system can rapidly respond to more quickly satisfy the user.  
\end{abstract}

\ccsdesc[300]{Information systems~Recommender systems}
\keywords{conversational recommendation, user simulation, evaluation methodology}


\maketitle

\section{Introduction}

\begin{figure}[ht!]
\centering
\includegraphics[height=5cm]{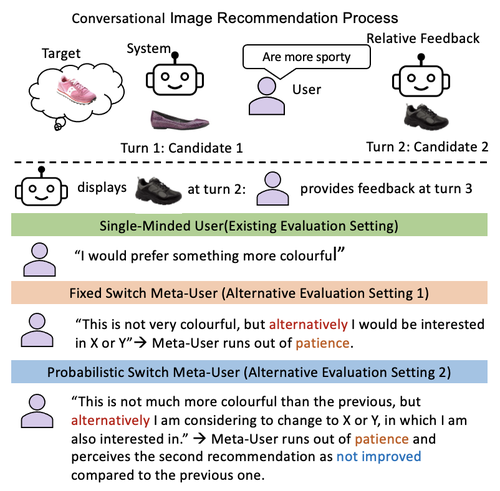}
\caption{Example of a Conversational Image Recommendation interaction (top).  Bottom: Illustration of the limitations of the evaluation setting; currently, users are assumed to have a single target in mind (green). Instead, the presence of alternatives in either a fixed (orange) or probabilistic way (blue) could more accurately represent a user need.}\label{fig:example_alt}\vspace{-\baselineskip}
\end{figure}

\looseness -1 In recent years, online shopping has become increasingly popular, leading to the development of e-shopping platforms, which help users with product search~\cite{rowley2000product,zou2019learning}, as well as a growing trend in research on fashion recommendation. As a result, various approaches and related datasets have been proposed that focus mainly on image recommendation and retrieval~\cite{han2017learning,he2016ups,liu2016deepfashion,mcauley2015image}. Due to the highly personalised nature of fashion recommendation, conversational recommendation systems (CRS) – which allow users to express their feedback as natural language~\cite{kovashka2013attribute,kovashka2017attributes,yu2017fine} – have gained importance in this domain.
In such settings, users’ preferences are collected over a number of interaction turns with the system. This task is typically described as Conversational Image Recommendation~\cite{guo2018dialog}, a sub-task of CRS, \mv{which is formulated as a reinforcement learning problem that optimises the rank of the target image item, where the user input has the form of natural language feedback, and the system output is a ranked list of image items. An example of a user interacting with such a CRS is shown in Figure~\ref{fig:example_alt} (top). Here, the user provides relative natural language feedback on a {\em candidate} item by mentally comparing its visual characteristics with their desired or {\em target} item at each turn~\cite{guo2018dialog,wu2021fashion,wu2021partially,wu2022multi,yu2019visual}.}

Since the training and evaluation of CRS uses a reinforcement learning approach~\cite{guo2018dialog}, it requires access to a large amount of data~\cite{shi2019build}. To compensate for the lack of human data, training and evaluation is often based on the interactions with a {\em simulated} user that has a single desired target item “in mind" and acts as a surrogate for human users. This is suggestive of a known-item type of task~\cite{broder2002taxonomy}, where the target item is specific and assumed to exist in the item catalogue. In our view, this negatively impacts the existing evaluation setting. The problem is depicted in Figure~\ref{fig:example_alt} (bottom). In particular, in existing user simulators, there is the option to request for one already determined target (green). Still, there is no option for a user to request one or more alternatives to their target item, either deterministically after a few turns (orange), or probabilistically after comparing the candidate to previously seen items (blue). However, these cases are possible in a real-life shopping scenario.

\pageenlarge{2} Indeed, the current fashion recommendation evaluation setting presents some limitations: First, the {\em realisticity} of an interaction does not aid user experience. Specifically, the simulated user is assumed to be infinitely patient, and willing to interact with the CRS for a large number of turns until the target item is found. This setting is not representative of a real user experience, where a user might become frustrated. To complement this, a simulated user is assumed to be single-minded, meaning that it is not flexible enough to change its strategy or initial plan. On the contrary, recommender systems are typically used to aid exploratory user behaviour~\cite{broder2002taxonomy,o2006race}, and therefore, by persisting on a single desired item, users are not exploring the product space. Moreover, unlike information retrieval systems, which are evaluated using test collections that aim to provide a reasonably {\em complete} coverage of relevant documents~\cite{craswell2020overview,dalton2020trec}, recommender systems suffer from a lack of completeness~\cite{chaney2018algorithmic,jadidinejad2020using,jadidinejad2021simpson}. In particular, search engines use {\em pooling} of documents retrieved from various systems and a per query {\em relevance judging} of pooled documents to obtain \cm{more complete assessments}~\cite{craswell2020overview,dalton2020trec}. For example, the MSMARCO test collection has thousands of queries containing shallow judgements, while the TREC Deep Learning track provided $\sim$100 queries with deeper judgements~\cite{craswell2020overview}. In this regard, it was found that the sparse MSMARCO assessments are not a suitable replacement for more complete assessments~\cite{macavaney2023one}. Similarly, the presence of more reliable relevance judgments for CRS would benefit the reliability of their evaluation. In this paper, we show that relevance judgments for CRS items can be obtained by directly asking users about alternative preferences, thus allowing the simulated user to update their preferences during the dialogue.

In short, this work contributes: (i) The first extended dataset for fashion recommendation that contains relevance assessments about the presence of sufficient alternatives for 200 known target items by real users for each of the different fashion categories (shoes and dresses) derived from two popular fashion CRS datasets, namely Shoes~\cite{berg2010automatic,guo2018dialog} and FashionIQ~\cite{wu2020fashion}.  (ii) two {\em meta-simulators} for users, which use these relevance judgments and wrap the existing user simulator to provide feedback for possible alternatives items; and (iii) a study of existing CRS models evaluated with and without alternatives. In our analyses, we find that the existing single-target evaluation of CRS underestimates their effectiveness, and when simulated users are allowed to instead consider alternatives, the system can rapidly respond to more quickly satisfy the user. We share our collected data and code at \url{https://github.com/mariavlachou/AlterEval_CRS/}. To facilitate future pre-registered experiments using the dataset, we also make it available at \url{https://osf.io/yueab/}.


\section{Related Work}\label{sec:rw}\pageenlarge{1}

User simulation has been used extensively in interactive systems~\cite{chung2004developing,griol2013automatic,owoicho2023exploiting,sun2023metaphorical,verberne2015user,zhang2020evaluating,zhang2022analyzing}. For example, improved performance was observed in conversational search systems by simulating user feedback~\cite{owoicho2023exploiting} or user satisfaction using annotated data~\cite{sun2021simulating}. As for CRS, user simulators are common practice for training and evaluation, since models use a {\em reinforcement learning (RL)} approach, where an action is taken based on estimated rewards~\cite{zhang2020evaluating,guo2018dialog,vakulenko2019qrfa,wu2021fashion} while optimising for long-term performance~\cite{shi2019build}. Training CRSs requires a large amount of data~\cite{li2016user,shi2019build}; to account for this, user simulators are used as a surrogate of human behaviour~\cite{li2016user,shi2019build}. Recent work builds on an agenda-based simulation framework that uses push-and-pull operations to update the user needs per turn~\cite{balog2021conversational,schatzmann2007agenda,vakulenko2019qrfa,zhang2020evaluating}. Regarding Conversational Image Recommendation~\cite{guo2018dialog,wu2022multi,wu2021fashion,wu2021partially,yu2020towards}, the state-of-the-art simulator framework is {\em relative captioning}, where the dialog system is trained on human-annotated captions that describe the visual differences between two image items. Other annotation-based CRS simulation approaches collect datasets limited to rating the level of dialogue success or user satisfaction~\cite{sun2021simulating}. On the other hand, our work extends the completeness of the ground truth by adding options to the target space. In other words, we enrich simulated users with a target {\em group} instead of single target items. In that sense, our approach is similar to~\cite{sun2023metaphorical}, who assume an analogical thinking of users, i.e., users comparing new items with prior knowledge. Still, they do not provide alternative preferences, which is the basis of our work.

In parallel, the evaluation of recommender systems lacks completeness, as past interactions are ``replayed'' and the ability of the recommender system to predict the hidden ``future'' interaction(s) is measured by classical evaluation measures such as MRR and NDCG. This tends to favour systems that behave similarly to the system originally deployed when the user interactions were collected~\cite{chaney2018algorithmic,jadidinejad2020using,jadidinejad2021simpson}. In contrast, search engine evaluation uses test collections~\cite{sanderson2010test}, which combine two techniques for obtaining a more {\em complete} coverage of relevant documents: the {\em pooling} of documents retrieved by a number of diverse effective systems, and the explicit judging of the relevance of all pooled documents to a user's query. Incomplete test collections are well known to result in unreliable evaluation~\cite{buckley2007bias,buckley2004retrieval}. Recently, Craswell et al.~\cite{craswell2020overview} found a good correlation between evaluation using thousands of single known relevant queries versus using deeply judged TREC queries; however, pseudo-relevance feedback techniques have been shown to work on the latter but not the former~\cite{wang2023colbert}. Pooling and assessing is typically not used for recommendations, as the user's exact information needs are not clear. However, for CRS with an assumed target item, we argue that it is possible to ask a 3rd party assessor to consider what other items they might prefer as relevant alternatives. In this way, we develop more complete test collections for fashion-based CRS (using alternative target items) and a more realistic user simulator that can use these alternatives during evaluation.

\section{Enriching of CRS Datasets with Alternatives}\label{sec:dataset}

We now describe how we enriched two fashion CRS datasets with alternative judgements to create our dataset of alternative judgements consisting of different fashion categories.

\subsection{Original Datasets}\label{ssec:data}
To build our dataset, we use two popular datasets in the conversational fashion recommendation domain, namely the Shoes~\cite{berg2010automatic,guo2018dialog} dataset and the FashionIQ Dresses dataset~\cite{wu2021fashion}. In particular, Shoes contains 4658 test target images, while Dresses contains 2454 test images. These datasets were originally collected to train and evaluate CRS in the relative captioning setting~\cite{guo2018dialog,wu2020fashion}. For this reason, each target image is accompanied by a corresponding paired candidate image, as well as a relative critique or caption per candidate-target pair, which describes the relative visual differences between the candidate and target image pairs and are used as training data for the user simulator. For our task, we obtain {\em labels of relevance} (whether a set of candidate images is a sufficient alternative to a given target image) for a portion of the target images in the original datasets, which we treat as different {\em fashion categories}. 
 

\subsection{User Study Details}\label{ssec:user_study}
Our user study can be further divided into two main stages: target pooling (described in section ~\ref{sssec:pooling}) and data collection (described in section ~\ref{sssec:data_collection}). In particular, we use Amazon Mechanical Turk\footnote{https://www.mturk.com} to obtain assessments on alternative items. Target pooling was completed in August 2023, while data collection extended from 18 August 2023 until 4 October 2023. It involved several smaller test batches for each fashion category before the final deployment of the image set. Next, we describe further how each of the two sub-tasks was conducted.

\subsubsection{Target Pooling}\label{sssec:pooling}
The purpose of the study was to collect relevance judgments for CRS systems to introduce a parallel to the test collection paradigm in information retrieval test collections. Therefore, we need to ensure that we select a number of representative target image items (as the queries in test collections) for which users will provide their opinions about relevant alternatives. This brings two requirements: (i) Select a sufficient number of items to achieve a level of generalisability. (ii) Select items with prior knowledge derived from different systems - this ensures a variety in the sample of items that are more representative of the underlying population of items. To account for the first requirement, we estimate the required number of sampled target images with a power analysis using the reported correlation values of a recent Conversational Performance Prediction analysis~\cite{vlachou2022performance} as effect sizes (by converting the correlation values to Cohen's $d$), a significance level (alpha value) of $\alpha = 0.05$, to achieve a power of 90\%. The power analysis estimation, which can be found in Table~\ref{tab:power_analysis}, provided a number less than 200 targets for each fashion category (shoes, dresses), but we opted for 200 from each to obtain a sufficient number of targets. 

\begin{table}[tb]
\caption{Summary of the required sample size of target image items from each original dataset (from power analysis).}\label{tab:power_analysis}
\begin{center}
\resizebox{0.6\linewidth}{!}{
\begin{tabular}{c|ccc}
\hline
Dataset & Spearman's $\rho$ & Cohen's $d$ & Required Sample \\ \hline
Shoes   & -0.423     & -0.933  & 54     \\
Dresses & -0.281     & -0.585  & 128    \\ \hline
\end{tabular}}
\end{center}
\end{table}

\begin{figure}[tb]
\centering
\includegraphics[width=85mm]{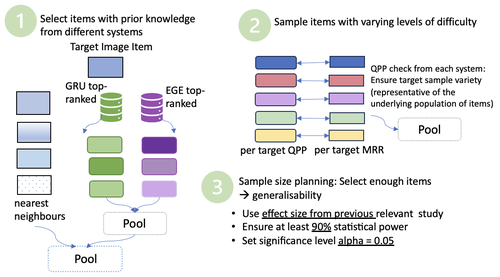}
\caption{Schematic representation of our target pooling strategy. We select candidate items from top-ranked items and nearest neighbours, and use targets with different levels of difficulty. Finally, we ensure a sufficient amount of relevance assessments by basing our estimation on a power analysis.}\label{fig:pool}\vspace{-\baselineskip}
\end{figure}

\looseness -1 To account for the second requirement, we select the target images by sampling 200 target items from each fashion category with varying levels of difficulty (we checked this by conducting a preliminary Query Performance Prediction~\cite{carmel2010estimating,cronen2002predicting} analysis of the sampled items using score-based predictors~\cite{roitman2017enhanced,shtok2012predicting}). In addition, for assessment, we derive a pool of candidate images for each target by using existing state-of-the-art CRS models for conversational image recommendation, specifically GRU~\cite{guo2018dialog,hidasi2015session} and EGE~\cite{wu2021partially} (detailed further in Section~\ref{ssec:systems} below). In particular, we select both the nearest neighbours (in their corresponding image embedding spaces) to the target (60\%) and their top-retrieved images of the final evaluation turn (40\%).
We place more importance on the nearest neighbours, because we are more interested in the similarity of the images rather than how each CRS model ranks them. Specifically, we use the top-4 ranked nearest neighbours of each target item from each CRS model, and the top-3 top-ranked results from each CRS model at turn 10. This results in a total of 14 candidate images per target item. To ensure no duplicates, we checked how many items overlap between the two CRS models (both for nearest neighbours and retrieved results), and in case of common entries, we replaced them with additional items from lower ranks. A schematic representation of our data pooling approach in steps is presented in Figure~\ref{fig:pool}. In summary, we followed a detailed strategy for data pooling, which resulted in an amount of precisely estimated and representative target items from each fashion category, which already exceed both the more recent TREC Deep Learning~\cite{craswell2020overview} test collection query sets by roughly four times and the TREC CAsT for conversational tasks~\cite{dalton2020cast,dalton2020trec}.

\subsubsection{Data Collection}\label{sssec:data_collection}
\begin{figure*}[ht!]
\centering
\includegraphics[height=7cm]{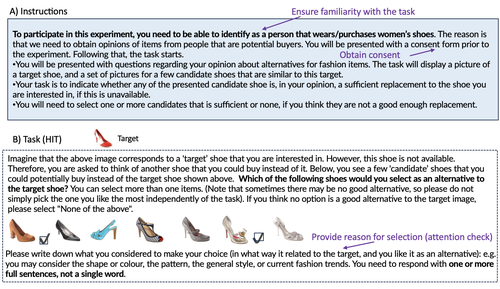}
\caption{Example HIT (Amazon Mechanical Turk task) from our user study for the Shoes dataset. The target item appears at the top, while the worker is instructed to select one or more alternatives from the items appearing below as candidates.}\label{fig:hit_example}
\end{figure*}

As mentioned, we conduct our study on Amazon Mechanical Turk, which has been used as a platform for data collection in several online studies for various conversational systems~\cite{jurcicek2011real,owoicho2023exploiting,sun2021simulating,sun2018conversational} and CRS systems~\cite{liu2020towards,zhang2020evaluating,zhou2020towards}, and also in our setting of interest, namely relative captioning, where the original datasets were obtained with crowd-sourcing~\cite{guo2018dialog,wu2020fashion}. We take some additional steps to ensure the representativeness and the knowledge level of our sample. Specifically, participants are selected based on their location (US, to ensure an adequate level of English) and to the extent they could identify with a person who wears dresses or women's shoes (familiarity with and knowledge of the target items), and are paid based on the rules of Mechanical Turk. \maria{We obtained institutional ethical approval for the study, and we paid participants \$0.63 for each MTurk task (or {\em HIT}) for a total duration of 3 minutes (this is above the living wage in our country), making a total cost of the study was \$305 (we rejected only 3 HITs for spammy behaviour)}. In our study, we simulate a real user (online) shopping scenario. Participants are instructed that each presented target image is an item they want to buy. Simultaneously, participants see a set of candidate image items that could be a sufficient alternative to the target, which, as instructed, is not available in the catalogue. The task is to select, out of the displayed set of candidates, the ones (if any) that best satisfy the user need as an alternative \cm{to the target item}. Finally, participants are asked to indicate the reason for their selection (this also works as an attention check, as we ask them to respond with a full sentence and set a minimum required length). Each participant was allowed to complete one or more HITS. For each of the resulting alternative labelled fashion categories of our dataset, there were on average 3.5 identified relevant alternatives per target image. We performed a second round of assessment on 40 target items (10\% from each dataset), and measured assessor agreement. We observed a Cohen's $\kappa$ between the two sets of judgments of 0.87, demonstrating a high level of agreement. The procedure of the user study is described in Figure~\ref{fig:hit_example}. Specifically, participants are shown the task instructions, and we ensure that they are familiar with the task while obtaining their consent. Then, the task starts, which asks them to select one or more alternatives as relevant items if they believe they match the above-presented assumed target.

\subsection{Inspection of the Collected Data}\label{ssec:qualitative}
\begin{figure*}[tb]
    \centering
     \begin{subfigure}[b]{\linewidth}
     \includegraphics[width=\textwidth,height=2cm]{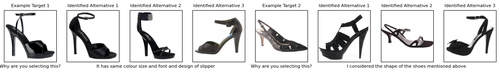}
     \caption{Shoes}
     \end{subfigure}
     \begin{subfigure}[b]{\linewidth}
     \includegraphics[width=\textwidth,height=2cm]{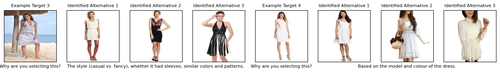}
     \caption{Dresses}
     \end{subfigure}
    \caption{Examples of identified alternative relevant items from our user study for Shoes (top) and Dresses (bottom). The justification of selection for the set of alternatives per item is provided as a caption.}
    \label{fig:items_example}
\end{figure*}

A visual inspection of the collected dataset is presented in Figure~\ref{fig:items_example}, where we see typical examples of original target image items from both fashion categories together with the selected alternative relevance judgments for each and the justification of selection by the participant. In general, the identified alternatives are well-aligned with their corresponding target, while many visual features are in common. As for the reasons for selection, some influential characteristics are the colour and pattern of the items, the size and shape, and their general style similarity. When inspecting the data qualitatively, we noticed that colour is mentioned frequently, followed by pattern and shape, while participants also indicated that the selected items also reflect their taste (as being similar to the target, which matches their preferences). The statistics of Fashion-AlterEval are summarised in Table~\ref{tab:stat} (the last column denotes the annotations collected for each presented alternative candidate of a given target as true or false).

\begin{table}[tb]
\caption{Statistics of the new dataset Fashion-AlterEval.}\label{tab:stat}\vspace{-0.5em}
\resizebox{0.8\linewidth}{!}{
\begin{tabular}{l|lllll}
\hline
\multicolumn{1}{c|}{Category} & n\_target & n\_assessed  & n\_relevant & avg relevant & n\_annotations per target \\ \hline
Shoes                        & 4658      & 200         & 190         & 3.5   & 14       \\
Dresses                      & 2454      & 200         & 181         & 3.5    & 14      \\ \hline
\end{tabular}}\vspace{-0.5em}
\end{table}

In the next section, we show how our alternatives-based relevance assessments can be used to test CRS under different user simulators - for example, simulated users that are allowed to change their preferences towards alternative items. Indeed, we propose new user simulators and use them to evaluate three representative fashion CRS systems using the new dataset.


\section{Simulated Users with Alternatives}\label{sec:simulator_desc}
We start by providing some general notation about the principles of a single-target simulator. A simulator using relative captioning is an object with a function that takes as input the user's id, the current top-ranked candidate image, $top\_ranked$, and the user's actual target image $target$. It then calls a learned relative captioning model, which has been trained given $target$ to critique $top1_k$, and return a text string describing the visual differences between  $top1_k$ and $target$:
\begin{equation}\label{eq_sim_1}
Usersim.critique(user, top1_k, target) \rightarrow \mathcal{T}
\end{equation}
Next, we describe the functionality and intuitions of our proposed meta user simulators that use either a fixed switch to alternatives (Section~\ref{ssec:sim_tol}) or a probabilistic switch based on the estimation of gains and losses (Section~\ref{ssec:sim_prob}). 
\subsection{Meta-Simulator with Fixed Alternative Selection}\label{ssec:sim_tol}
Moving forward, for a user that considers alternative items, our intuitions are: (I1) their patience (tolerance) critiquing for a single target item may run out after number of turns, (I2) when selecting an alternative as a new target, they are influenced by the current item they see, and (I3) the existing relative captioner-based user simulator can be called with a new target. Indeed, intuitions I1-I3 represent an average user profile or a user that opts for a simple strategy in the lack of time and resources (or a less selective user).  
In particular, Algorithm~\ref{alg:next_k} provides pseudocode for our tolerance-based meta-user simulator procedure. Specifically, $MetaSimTol$ first requires knowledge of all possible alternative items for target items. This is akin to the ``qrels'' in test collection-based evaluation (i.e., in a TREC test collection, queries are accompanied by the relevance of documents, which is similar to our relevance judgments). Then, when the meta-user simulator is called, if the turn number exceeds the patience $tolerance$ parameter, alternatives are considered (line 2, addressing intuition I1); Among all of the alternatives for a given target, we select the alternative that is closest in image similarity to the current top-ranked image as the target (line 3, addressing I2). The existing relative captioning-based user simulator is then asked to critique the newly selected target (line 6, I3). Note that we choose to consider the target as part of the alternatives, so that the ranker can choose between the nearest item at a later stage. Using the new meta-simulator algorithm, the updated learned relative captioning model (which is still a critiquing method) can be written as:
\begin{equation}\label{eq_sim_2}
Metausersim.critique(user, top1_k, all\_targets) \rightarrow \mathcal{T}
\end{equation}

\subsection{Meta-Simulator with Probabilistic Gain-Loss Alternative Selection}\label{ssec:sim_prob}
\begin{figure}[tb]
\centering
\includegraphics[height=4cm]{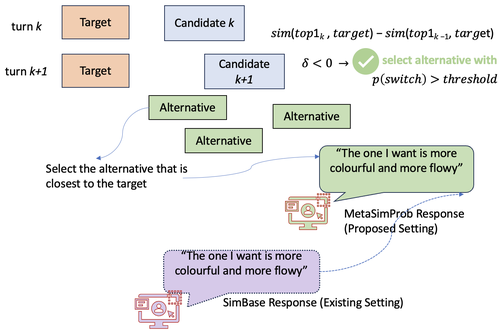}
\caption{Schematic representation of our $MetaSimProb$ meta-simulator. If the difference of target-candidate similarity between turns is negative, the meta-user opts for a switch to an alternative with probability $P_{switch}.$}\label{fig:meta_sim}\vspace{-\baselineskip}
\end{figure}

\looseness -1 The proposed meta-simulator above offers the possibility to select one or more alternatives at different points of a conversation. However, with increasing personal involvement (the more they are interested in finding something), further specifications are added to the selection. In this regard, we are inspired by the {\em gain-loss framing effect}~\cite{tversky1981framing}, a heuristic where people choose a certain option when a decision is framed as a gain, while they tend to prefer a risky option when it is presented as a loss. Indeed, changing the focus from losses to gains decreases participants’ willingness to take risks when asked to select an action to combat a dangerous disease (gamble to get a better outcome rather than take a guaranteed result)~\cite{tversky1981framing}. This effect was replicated in more recent studies with larger sample sizes~\cite{chick2016framing,klein2014investigating,pinon2005meta}. In our case, a user in a shopping scenario can be seen as someone who calculates gains and losses. During a dialogue, a user provides a critique to a candidate by comparing it with the candidate of the previous turn. If it is worse, they perceive it as a loss; if it seems more relevant, it is a perceived gain. This is an added intuition for an involved meta-user (I4).
In a gain, the user’s probability of selecting an alternative is reduced, while after a loss, the user risks getting an even more irrelevant item. Algorithm~\ref{alg:next_k_prob} provides pseudocode for our probabilistic gain-loss-based meta-user simulator procedure. Specifically, $MetaSimProb$ uses the same patience tolerance limit as Algorithm~\ref{alg:next_k} in line 2, but introduces I4 in line 3 by comparing the current candidate with the previous turn candidate. If this difference is negative (perceived loss), the meta-simulator selects an alternative with probability equal to a threshold (additional parameter $P_{switch}$) (I5, line 4). The process of our $MetaSimProb$  is shown in Figure~\ref{fig:meta_sim}. Note that we can obtain $MetaSimTol$ from this if we omit the step that includes $\delta$.

\begin{algorithm}
\caption{MetaSimTol}
\label{alg:next_k}
\begin{algorithmic}[1]
\Procedure{MetaUserSim.critique}{$k, top1_k$, $target$}

        \If{$turn > tolerance$}
            \State $target = alts[\argmax(sims(alts, top1_k))]$
            \State $switch = 1$
        \EndIf
        
        \State \textbf{return} Usersim.critique($turn, top\_ranked, target$)
\EndProcedure
\end{algorithmic}
\end{algorithm}

  

  

\begin{algorithm}
\caption{MetaSimProb}
\label{alg:next_k_prob}
\begin{algorithmic}[1]
    \Procedure{MetaUserSim.critique}{$k, top1_k, top1_{k-1}, target$}
        \If{$turn > tolerance$}
            \State $\delta = sim(top1_k, target) - sim(top1_{k-1}, target)$
            \If{$\delta < 0$ and $random() < P_{switch}$ }
            
                \State $target = alts[\argmax(sims(alts, top1_k))]$
                \EndIf 
        \EndIf    
        \State \textbf{return} Usersim.critique($k, top1_k, target$)
\EndProcedure
\end{algorithmic}
\end{algorithm}


\section{Experiments}\label{sec:results}

Having inspected our data qualitatively and after presenting our meta-simulators, we experiment quantitatively to address the following research questions:

{\bf RQ1} What is the impact of using: (a) an alternative-based meta-simulator and (b) a probabilistic gain-loss meta-simulator on the evaluation of existing CRS models? 

{\bf RQ2} What is the impact of patience and threshold of an alternative-based user simulator?

{\bf RQ3} Does the number of target image items affect the performance of CRS models?

\subsection{Setup: CRS models and User Simulators}\label{ssec:systems}
We deploy three CRS models using both the original relative captioning single-target evaluation setting and our own meta-simulator with alternatives. Each system retrieves 100 top items per turn.
\begin{itemize}
  \item A GRU model~\cite{guo2018dialog,hidasi2015session} with reinforcement learning (GRU-RL), which combines input representation with the historical information representation from the previous turn, and produces an updated aggregated representation vector. This is achieved with a gated recurrent unit (GRU)~\cite{cho2014learning} applied to its State Tracker. The updated historical information representation based on information from the current dialog turn. This formulation of the State Tracker leads to a memory-based design that sequentially takes into account information from user feedback to identify new candidate items~\cite{guo2018dialog}, and is, therefore, optimised for maximising short-term rewards.
  \item A GRU variant trained with supervised learning (GRU-SL), i.e.\ lacking short-term rewards during training.
  \item The Estimator - Generator - Evaluator (EGE) model~\cite{wu2021partially}, which considers interactive recommendation as a partially observable Markov decision process. In particular, EGE learns a policy that depends on observations but also on action histories (historical feedback and recommendations), and conditions its actions on the entire history. For that reason, and compared to GRU~\cite{cho2014learning,guo2018dialog}, it maximises longer-term rewards.
\end{itemize}
For each, we use the original relative captioning single-target simulator (denoted $SimBase$), and our meta-simulators with alternatives ($MetaSimTol$ and $MetaSimProb$). We retain the original training for these models using a user simulator that considers a single target item. For each meta-simulator, we use tolerance levels from turns 1 to 4. For $MetaSimProb$, we also vary the threshold (probability of switch to an alternative) as follows: We derive the probability of selecting the risky option in a perceived loss from the multi-site study of Klein et al.~\cite{klein2014investigating}, which is the replication attempt with the largest sample size. We convert Cohen's d to probability, which gives a value of 0.75. We, therefore, use a grid with increasing probabilities by 0.1 in both directions [0.55, 0.65, 0.75, 0.85, 0.95].

\subsection{Setup: Evaluation Measures}\label{ssec:measures}
\looseness -1 Following existing work in CRS, we use classical IR evaluation measures to evaluate the ability of each CRS system to retrieve relevant items. In particular, we measure the ability of the CRS to show the user's desired target at rank 1 (Success Rate @ 1) at each turn of the conversation. Moreover, as there is a ranking of images created at each turn, we use nDCG@10 and MRR@10 to evaluate the presence of target items in the ranking. Following \cite{guo2018dialog,wu2021partially}, we terminate conversations at turn 10; if a target item has been found before turn 1, then all evaluation measures are considered to be equal to 1. Finally, and differing from previous work, we consider the alternatives as relevant items for evaluation - in this case, a conversation is successful if any alternative (or the original target) is retrieved (this holds for evaluating the meta-simulator; we still use the classical evaluation setup for the base simulator). Measuring systems in different evaluation setups can identify how much existing CRS models can respond to changes in the user.

\subsection{RQ1 - Impact of alternative-based user simulator on the evaluation of CRS models}\label{ssec:impact_sim}

\begin{table*}[tb]
\caption{CRS model performance after applying our meta-simulators. The percentage of improvement at the last row indicates the difference in the quality of quantifying a user need (reflected in the system's performance) between $SimBase$ and any of the $MetaSimTol$ or $MetaSimProb$ specifications in the same group of rows for the corresponding dataset.}\label{tab:results_all}
\resizebox{0.8\textwidth}{!}{
\begin{tabular}{@{}lcccccccccccc@{}}
\toprule
                                                                                                        & \multicolumn{6}{c}{Shoes}                                                                    & \multicolumn{6}{c}{Dresses}                                                       \\ \midrule
\multicolumn{1}{l|}{\multirow{2}{*}{\begin{tabular}[c]{@{}l@{}}CRS Model \\ \& Simulator\end{tabular}}} & \multicolumn{2}{c}{GRU-SL} & \multicolumn{2}{c}{GRU-RL} & \multicolumn{2}{c|}{EGE}           & \multicolumn{2}{c}{GRU-SL} & \multicolumn{2}{c}{GRU-RL} & \multicolumn{2}{c}{EGE} \\ \cmidrule(l){2-13} 
\multicolumn{1}{l|}{}                                                                                   & NDCG@10         & MRR@10         & NDCG@10         & MRR@10         & NDCG@10  & \multicolumn{1}{c|}{MRR@10}   & NDCG@10         & MRR@10         & NDCG@10         & MRR@10         & NDCG@10       & MRR@10        \\ \midrule
\multicolumn{1}{l|}{SimBase}                                                                            & 0.209        & 0.196       & \textbf{0.303}        & 0.291       & 0.277 & \multicolumn{1}{c|}{0.263} & 0.072        & 0.068       & 0.099        & 0.269       & 0.085      & 0.084      \\
\multicolumn{1}{l|}{MetaSimTol-tol1}                                                                    & 0.237        & 0.521       & 0.211        & 0.519       & 0.288 & \multicolumn{1}{c|}{0.593} & 0.114        & 0.321       & 0.104        & 0.281       & 0.194      & 0.475      \\
\multicolumn{1}{l|}{MetaSimProb-tol1}                                                                   & 0.240        & 0.530       & 0.246        & 0.535       & 0.293 & \multicolumn{1}{c|}{0.602} & 0.121        & 0.365       & 0.102        & 0.315       & 0.237      & 0.520      \\
\multicolumn{1}{l|}{MetaSimTol-tol2}                                                                    & 0.237        & 0.494       & 0.230        & 0.543       & 0.286 & \multicolumn{1}{c|}{0.611} & 0.125        & 0.353       & 0.099        & 0.269       & 0.225      & 0.541      \\
\multicolumn{1}{l|}{MetaSimProb-tol2}                                                                   & 0.228        & 0.536       & 0.252        & 0.570       & 0.326 & \multicolumn{1}{c|}{0.640} & \textbf{0.139}        & \textbf{0.388}       & 0.108        & 0.310       & 0.242      & 0.559      \\
\multicolumn{1}{l|}{MetaSimTol-tol3}                                                                    & 0.245        & 0.516       & 0.239        & 0.556       & 0.300 & \multicolumn{1}{c|}{0.589} & 0.126        & 0.378       & 0.108        & 0.299       & 0.232      & 0.558      \\
\multicolumn{1}{l|}{MetaSimProb-tol3}                                                                   & 0.240        & \textbf{0.545}       & 0.264        & \textbf{0.573}       & 0.328 & \multicolumn{1}{c|}{0.657} & 0.132        & 0.381       & \textbf{0.121}        & 0.332       & 0.242      & 0.563      \\
\multicolumn{1}{l|}{MetaSimTol-tol4}                                                                    & \textbf{0.248}        & 0.511       & 0.239        & 0.543       & \textbf{0.332} & \multicolumn{1}{c|}{0.655} & 0.134        & 0.368       & 0.105        & 0.299       & \textbf{0.258}      & \textbf{0.587 }     \\
\multicolumn{1}{l|}{MetaSimProb-tol4}                                                                                        & 0.242        & 0.538       & 0.245        & 0.558       & 0.329 & \multicolumn{1}{c|}{\textbf{0.659}}                      & 0.136        & 0.381       & 0.114        & \textbf{0.333}       & 0.252      & 0.572      \\
\multicolumn{1}{l|}{\% Improv.} & 17.06 & 94.19 & -13.75 & 64.80 & 18.06 & \multicolumn{1}{l|}{85.90} & 63.50 & 140.35 & 20 & 21.26 & 100.87 & 149.92
\\ \bottomrule
\end{tabular}}
\end{table*}

Table~\ref{tab:results_all} shows the performance of the three CRS models at the final turn of a conversation before and after applying our alternative-based meta-simulators on Shoes and Dresses. Each group of columns indicates a separate CRS model, and within each, we include the different versions of $MetaSimTol$ and $MetaSimProb$ for the different tolerance levels (for example, MetaSimProb-tol1 denotes a gain-loss meta-simulator with tolerance at turn 1), thus allowing the comparison of the two meta-simulators for the same tolerance level. For each tolerance level of $MetaSimProb$, we report the metric value of the highest performing threshold level, as described in Section~\ref{ssec:measures}. To answer RQ1a), we consider the extent to which moving from no alternatives to alternatives included in the user simulator (any type) can have an impact on performance. Therefore, at the end rows, we include the percentage of improvement of any of the $MetaSimTol$ or $MetaSimProb$ specifications compared to the traditional setting with the non-alternative user simulator ($SimBase$) for the particular CRS model-dataset combination.

Overall, we observe improved estimation of user needs (reflected in how well we estimate performance) on both evaluation metrics and both Shoes and Dresses. For Shoes, there are considerable improvements on MRR@10 across all three CRS models. For instance, the highest improvement in MRR@10 is observed for the GRU-SL model, followed by EGE with only small differences. Surprisingly, for NDCG@10,  we observe negative values for GRU-RL, which means that NDCG@10 estimation of performance does not improve or even drops when introducing alternatives for this model. Improvements are in general, greater for MRR@10 than for NDCG@10. As for Dresses, we observe improvements across both evaluation metrics, especially EGE on both metrics and GRU-SL MRR@10. Unlike Shoes, for Dresses we observe a positive difference in all cases, and specifically for MRR@10, our estimation is markedly increased when adding the alternative options setting. To answer RQ1a), the impact of using an alternative-based simulator is marked positive when evaluating existing CRS models. This suggests that the previous single-target-based user simulators were underestimating the effectiveness of the CRS for real users.

To answer RQ1b), we need to compare between $MetaSimTol$ and $MetaSimProb$ for each level of tolerance within each model and fashion category. For tolerance at turn 4, we see that the two simulators converge across models and only differ slightly at turn 10 in favour of one or the other. As for tolerance of 1, NDCG@10 does not improve in general, but MRR@10 increases towards turn 10. Finally, we note that for tolerance levels 2 and 3, $MetaSimProb$ is consistently better than $MetaSimTol$ except for a few cases in NDCG@10 Shoes. Overall, introducing $MetaSimProb$ shows improved estimation of a user need (compared to $MetaSimTol$), less for NDCG@10 and more for MRR, and is more evident for the two GRU variants than EGE. This can be explained by the fact that GRU mainly considers information from the previous turn, and is therefore, more aligned with the intuition of $MetaSimProb$, which compares the similarity of targets in a pairwise manner (from one turn to the next). In contrast, EGE considers the entire dialogue history, and the results indicate that using $MetaSimTol$, the models can perform sufficiently well. Note that despite this observation, the two CRS models contributed equally in the data pooling process and therefore, the resulting judgments are not biased. To answer RQ1b), in the majority of cases, $MetaSimProb$ is higher, and therefore, using a gain-loss simulator adds value.

Note that typically, we do not compare performance over different "qrels", since, due to the increased number of relevant items, we should expect higher performances on datasets with larger numbers of relevant items. Still, we measure the estimation of performance (how accurately we predict effectiveness with either one or more relevant items) and do not compare the performance itself. Therefore, we opt for this analysis, as it shows us how changing the user simulator (the evaluation setting and relevance judgments) can increase estimation of a user need on the same model, which also reflects the usefulness of our collected relevance judgments.

\subsection{RQ2 -  Impact of patience and threshold}

\begin{figure}[tb]
    \centering
     \begin{subfigure}[b]{0.20\textwidth}
     \includegraphics[width=\textwidth]{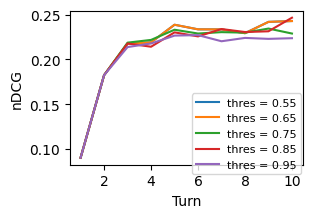}
     \caption{Shoes tol=1}
     \end{subfigure}
     \begin{subfigure}[b]{0.20\textwidth}
     \includegraphics[width=\textwidth]{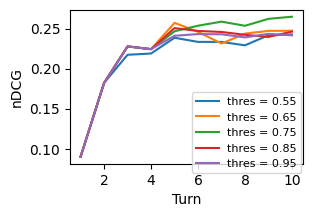}
     \caption{Shoes tol=3}
     \end{subfigure}
     \begin{subfigure}[b]{0.20\textwidth}
     \includegraphics[width=\textwidth]{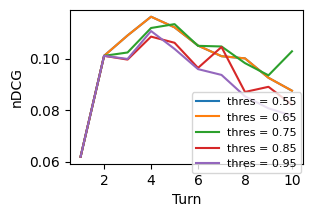}
     \caption{Dresses tol=1}
     \end{subfigure}
     \begin{subfigure}[b]{0.20\textwidth}
     \includegraphics[width=\textwidth]{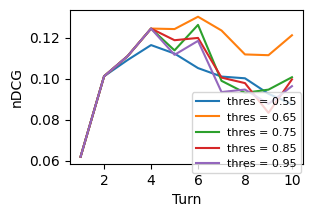}
     \caption{Dresses tol=3}
     \end{subfigure}
    \caption{NDCG@10 performance for the GRU-RL model for tolerance 1 and 3 Shoes and Dresses using our $MetaSimProb$.}
    \label{fig:plots_gru}
\end{figure}


To answer RQ2, we examine how CRS performance (when using $MetaSimProb$ to generate feedback) changes with varying threshold values for different tolerance levels. Figure~\ref{fig:plots_gru} shows the NDCG@10 performance at tolerance levels 1 and 3 for Shoes and Dresses for the GRU-RL model. Each line corresponds to a different threshold level. First, we observe that changing the tolerance level does not change the pattern of performance within each fashion category (performance estimation is similar for each category across patience levels). Second, we notice that the different threshold levels of Shoes roughly converge across turns. In contrast, Dresses presents higher variations among threshold levels. Finally, while for EGE tolerance has the effect of converging the threshold levels (as we discovered in a manual inspection), in GRU-SL, performance varies more intensely with increasing patience. This is the main identified interplay between tolerance, threshold, and CRS model.

\begin{figure}[ht!]
    \centering
     \begin{subfigure}[b]{0.40\linewidth}
     \includegraphics[width=\textwidth]{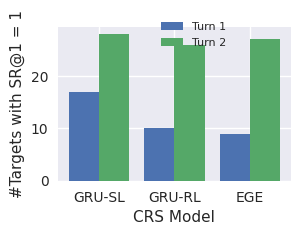}
     \caption{Tolerance = 1}
     \end{subfigure}
     \begin{subfigure}[b]{0.40\linewidth}
     \includegraphics[width=\textwidth]{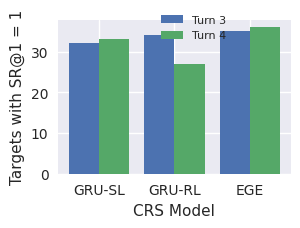}
     \caption{Tolerance = 3}
     \end{subfigure}
    \caption{\looseness -1 SR for Shoes before and after selecting an alternative with $MetaSimTol$ either after turn 1 or after turn 3.}
    \label{fig:sr_tol}\vspace{-1em}
\end{figure}

Another notable observation comes from inspecting the SR of Shoes when comparing before and after selecting an alternative with $MetaSimTol$ either early or later. Using the simpler meta-simulator, we notice that the earlier a user selects an alternative, the more intense we see the difference in representing the user need (as reflected in how well the CRS identifies a target). On the other hand, this also shows that when being more patient, the system is already doing better (than tolerance at turn 1) at the turn before the alternative selection for all models.

\begin{figure}[ht!]
    \centering
     \begin{subfigure}[b]{0.40\linewidth}
     \includegraphics[width=\textwidth]{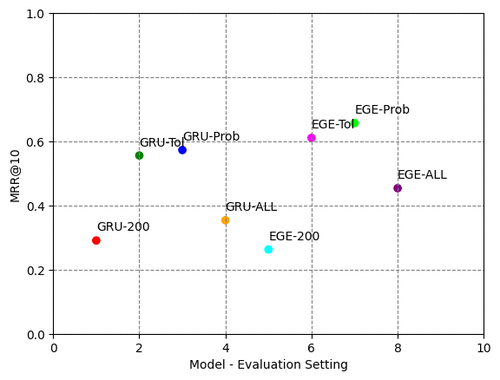}
     \caption{Shoes}
     \end{subfigure}
     \begin{subfigure}[b]{0.40\linewidth}
     \includegraphics[width=\textwidth]{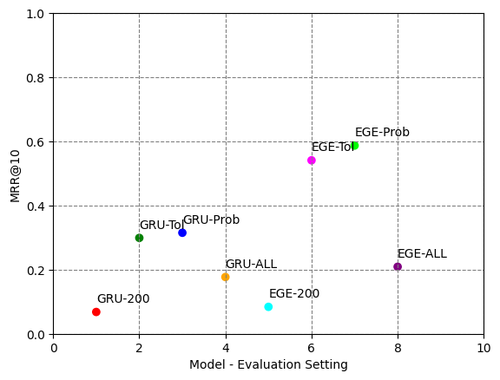}
     \caption{Dresses}
     \end{subfigure}
    \caption{\looseness -1 Inspection of the different evaluation settings (and corresponding user simulators) for GRU-RL and EGE and how this is reflected in MRR.}
    \label{fig:performance_eva}
\end{figure}

\subsection{RQ3 -  Number of target items to train a CRS}
Figure~\ref{fig:performance_eva} shows the MRR@10 for the GRU-RL and EGE models before and after introducing our meta-simulators. In this comparison, we include the performance obtained when training the same CRS models with the previous non-alternative evaluation setting but using all target items instead of a sample of 200. For both fashion categories, we notice that using our new evaluation setting, including alternative relevance judgments for 200 targets, results in higher MRR@10 than using thousands of targets, but with a reduced relevant item space. This finding supports our view of introducing a different evaluation setup and also links to findings from recent IR research, where the TREC Deep Learning track consists of around 100 queries with deeper judgments compared to the MSMARCO test collection, which contains thousands of shallow judgments~\cite{craswell2020overview}. This indicates that our collected dataset provides encouraging evidence towards this line of research.

\section{Conclusions}\label{sec:conc}
In this paper, we addressed the issue of obtaining relevance judgments in Conversational Recommendation Systems to achieve a more realistic recommendation setting and more accurately predict user preferences. In particular, we introduced a new relevance annotation approach that is based on directly asking real users about the relevance of items with respect to their similarity with a given target item. For this purpose, we conducted a user study that used crowd-sourcing to expand the existing well-used Shoes and FashionIQ Dresses datasets into a unified dataset, i.e., Fashion-AlterEval. In this way, we managed to extend the target space of a simulated user in the Conversational Image Recommendation setting by including the identified alternatives into the input datasets used to train the user simulator. In this regard, we showed how a sufficient amount of target items can be identified based on precise estimations that include pooling from diverse systems and various levels of difficulty, moving away from a perspective that selects a dataset size with a fixed required items. As a result, we ended up with an equivalent of TREC collections in information retrieval. Consequently, we created a more realistic novel dialog-based recommendation scenario, where a user is assumed to have a more widely defined information need, is flexible to adjust their strategy during a conversation according to what they see, and has the opportunity to change their mind. This was done by introducing two novel meta-user simulators that use the alternative relevant items for training and evaluation of CRSs. Our simulators inform the existing base (non-alternative) user simulator with knowledge of the alternative options to given target items, and therefore, allow the (simulated) users to change their mind during the CRS interaction. 

Overall, using our new test collection with alternative relevance judgments for evaluation, we found that the way a system estimates a user's need (as reflected in the CRS performance) is improved when changing the way a user simulator requests a given item. For the same CRS models, using these extended datasets and the corresponding meta-simulator for evaluation, we showed that previous (single-target) evaluations may underestimate the effectiveness of CRS systems on these datasets. Indeed, if they accept other alternative items and are willing to switch strategy, the system may satisfy them sooner. One limitation of our simulators is that they do not consider multiple alternatives. We leave this for future work, with the possibility of users ranking them in a study. As for Fashion-AlterEval, its use is not restricted to multi-turn recommendation settings. It could also be used for single-turn image retrieval, a concept more similar to traditional image retrieval test collections. Additionally, it could be used for different recommendation settings by modifying the meta-simulator accordingly.

\begin{acks}
This work was supported by the UKRI Centre for Doctoral Training in Socially Intelligent Artificial Agents, Grant number EP/S02266X/1.
\end{acks}


\begin{thebibliography}{8}
\bibitem{balog2021conversational}Balog, K. Conversational AI from an information retrieval perspective: Remaining challenges and a case for user simulation. {\em DESIRES}. (2021)
\bibitem{berg2010automatic}Berg, T., Berg, A. \& Shih, J. Automatic attribute discovery and characterization from noisy web data. {\em Proc. ECCV}. pp. 663-676 (2010)
\bibitem{broder2002taxonomy}Broder, A. A taxonomy of web search. {\em ACM Sigir Forum}. \textbf{36}, 3-10 (2002)
\bibitem{buckley2007bias}Buckley, C., Dimmick, D., Soboroff, I. \& Voorhees, E. Bias and the limits of pooling for large collections. {\em Information Retrieval}. \textbf{10}, 491-508 (2007)
\bibitem{buckley2004retrieval}Buckley, C. \& Voorhees, E. Retrieval evaluation with incomplete information. {\em Proceedings Of The 27th Annual International ACM SIGIR Conference On Research And Development In Information Retrieval}. pp. 25-32 (2004)
\bibitem{carmel2010estimating}Carmel, D. \& Yom-Tov, E. Estimating the query difficulty for information retrieval. (Morgan \& Claypool Publishers,2010)
\bibitem{chaney2018algorithmic}Chaney, A., Stewart, B. \& Engelhardt, B. How algorithmic confounding in recommendation systems increases homogeneity and decreases utility. {\em Proceedings Of The 12th ACM Conference On Recommender Systems}. pp. 224-232 (2018)
\bibitem{chick2016framing}Chick, C., Reyna, V. \& Corbin, J. Framing effects are robust to linguistic disambiguation: A critical test of contemporary theory.. {\em Journal Of Experimental Psychology: Learning, Memory, And Cognition}. \textbf{42}, 238 (2016)
\bibitem{cho2014learning}Cho, K., Van Merriënboer, B., Gulcehre, C., Bahdanau, D., Bougares, F., Schwenk, H. \& Bengio, Y. Learning phrase representations using RNN encoder-decoder for statistical machine translation. {\em ArXiv Preprint ArXiv:1406.1078}. (2014)
\bibitem{chung2004developing}Chung, G. Developing a flexible spoken dialog system using simulation. {\em Proc. ACL}. (2004)
\bibitem{craswell2020overview}Craswell, N., Mitra, B., Yilmaz, E., Campos, D. \& Voorhees, E. Overview of the TREC 2019 deep learning track. {\em ArXiv Preprint ArXiv:2003.07820}. (2020)
\bibitem{cronen2002predicting}Cronen-Townsend, S., Zhou, Y. \& Croft, W. Predicting query performance. {\em Proc. SIGIR}. (2002)
\bibitem{dalton2020cast}Dalton, J., Xiong, C., Kumar, V. \& Callan, J. Cast-19: A dataset for conversational information seeking. {\em Proc. SIGIR}. (2020)
\bibitem{dalton2020trec}Dalton, J., Xiong, C. \& Callan, J. TREC CAsT 2019: The conversational assistance track overview. {\em ArXiv Preprint ArXiv:2003.13624}. (2020)
\bibitem{griol2013automatic}Griol, D., Carbó, J. \& Molina, J. An automatic dialog simulation technique to develop and evaluate interactive conversational agents. {\em Applied Artificial Intelligence}. \textbf{27}, 759-780 (2013)
\bibitem{guo2018dialog}Guo, X., Wu, H., Cheng, Y., Rennie, S., Tesauro, G. \& Feris, R. Dialog-based interactive image retrieval. {\em Proc. NeurIPS}. pp. 678-688 (2018)
\bibitem{han2017learning}Han, X., Wu, Z., Jiang, Y. \& Davis, L. Learning fashion compatibility with bidirectional LSTMs. {\em Proc. ACM Multimedia}. (2017)
\bibitem{he2016ups}He, R. \& McAuley, J. Ups and downs: Modeling the visual evolution of fashion trends with one-class collaborative filtering. {\em Proc. WWW}. (2016)
\bibitem{hidasi2015session}Hidasi, B., Karatzoglou, A., Baltrunas, L. \& Tikk, D. Session-based recommendations with recurrent neural networks. {\em ArXiv Preprint ArXiv:1511.06939}. (2015)
\bibitem{jadidinejad2020using}Jadidinejad, A., Macdonald, C. \& Ounis, I. Using exploration to alleviate closed loop effects in recommender systems. {\em Proceedings Of The 43rd International ACM SIGIR Conference On Research And Development In Information Retrieval}. pp. 2025-2028 (2020)
\bibitem{jadidinejad2021simpson}Jadidinejad, A., Macdonald, C. \& Ounis, I. The simpson’s paradox in the offline evaluation of recommendation systems. {\em ACM Transactions On Information Systems (TOIS)}. \textbf{40}, 1-22 (2021)
\bibitem{jurcicek2011real}Jurcıcek, F., Keizer, S., Gašic, M., Mairesse, F., Thomson, B., Yu, K. \& Young, S. Real user evaluation of spoken dialogue systems using Amazon Mechanical Turk. {\em Proc. INTERSPEECH}. \textbf{11} (2011)
\bibitem{klein2014investigating}Klein, R., Ratliff, K., Vianello, M., Adams Jr, R., Bahn\'ik, S., Bernstein, M. \& Others. Investigating variation in replicability: a “many labs” replication project. Open Science Framework. (2014)
\bibitem{kovashka2013attribute}Kovashka, A. \& Grauman, K. Attribute pivots for guiding relevance feedback in image search. {\em Proc. ICCV}. (2013)
\bibitem{kovashka2017attributes}Kovashka, A. \& Grauman, K. Attributes for image retrieval. {\em Visual Attributes}. pp. 89-117 (2017)
\bibitem{li2016user}Li, X., Lipton, Z., Dhingra, B., Li, L., Gao, J. \& Chen, Y. A user simulator for task-completion dialogues. {\em ArXiv Preprint ArXiv:1612.05688}. (2016)
\bibitem{liu2016deepfashion}Liu, Z., Luo, P., Qiu, S., Wang, X. \& Tang, X. Deepfashion: Powering robust clothes recognition and retrieval with rich annotations. {\em Proc. CVPR}. (2016)
\bibitem{liu2020towards}Liu, Z., Wang, H., Niu, Z., Wu, H., Che, W. \& Liu, T. Towards Conversational Recommendation over Multi-Type Dialogs. {\em Proc. ACL}. (2020)
\bibitem{macavaney2023one}MacAvaney, S. \& Soldaini, L. One-Shot Labeling for Automatic Relevance Estimation. {\em ArXiv Preprint ArXiv:2302.11266}. (2023)
\bibitem{mcauley2015image}McAuley, J., Targett, C., Shi, Q. \& Van Den Hengel, A. Image-based recommendations on styles and substitutes. {\em Proc. SIGIR}. (2015)
\bibitem{owoicho2023exploiting}Owoicho, P., Sekulic, I., Aliannejadi, M., Dalton, J. \& Crestani, F. Exploiting Simulated User Feedback for Conversational Search: Ranking, Rewriting, and Beyond. {\em Proc. SIGIR}. pp. 632-642 (2023)
\bibitem{o2006race}O’Brien, J. The race to create a ‘smart’google. {\em Fortune Magazine}. (2006)
\bibitem{pinon2005meta}Piñon, A. \& Gambara, H. A meta-analytic review of framing effect: risky, attribute and goal framing. {\em Psicothema}. \textbf{17}, 325-331 (2005)
\bibitem{roitman2017enhanced}Roitman, H., Erera, S., Sar-Shalom, O. \& Weiner, B. Enhanced mean retrieval score estimation for query performance prediction. {\em Proc. ICTIR}. (2017)
\bibitem{rowley2000product}Rowley, J. Product search in e-shopping: a review and research propositions. {\em Journal Of Consumer Marketing}. \textbf{17}, 20-35 (2000)
\bibitem{sanderson2010test}Sanderson, M. \& Others Test collection based evaluation of information retrieval systems. {\em Foundations And Trends® In Information Retrieval}. \textbf{4}, 247-375 (2010)
\bibitem{schatzmann2007agenda}Schatzmann, J., Thomson, B., Weilhammer, K., Ye, H. \& Young, S. Agenda-based user simulation for bootstrapping a POMDP dialogue system. {\em NAACL-Short}. pp. 149-152 (2007)
\bibitem{shi2019build}Shi, W., Qian, K., Wang, X. \& Yu, Z. How to build user simulators to train rl-based dialog systems. {\em ArXiv Preprint ArXiv:1909.01388}. (2019)
\bibitem{shtok2012predicting}Shtok, A., Kurland, O., Carmel, D., Raiber, F. \& Markovits, G. Predicting query performance by query-drift estimation. {\em ACM Transactions On Information Systems (TOIS)}. \textbf{30}, 1-35 (2012)
\bibitem{sun2023metaphorical}Sun, W., Guo, S., Zhang, S., Ren, P., Chen, Z., Rijke, M. \& Ren, Z. Metaphorical User Simulators for Evaluating Task-oriented Dialogue Systems. {\em ACM Transactions On Information Systems}. (2023)
\bibitem{sun2021simulating}Sun, W., Zhang, S., Balog, K., Ren, Z., Ren, P., Chen, Z. \& Rijke, M. Simulating user satisfaction for the evaluation of task-oriented dialogue systems. {\em Proc. SIGIR}. pp. 2499-2506 (2021)
\bibitem{sun2018conversational}Sun, Y. \& Zhang, Y. Conversational recommender system. {\em Proc. SIGIR}. (2018)
\bibitem{tversky1981framing}Tversky, A. \& Kahneman, D. The framing of decisions and the psychology of choice. {\em Science}. \textbf{211}, 453-458 (1981)
\bibitem{vakulenko2019qrfa}Vakulenko, S., Revoredo, K., Di Ciccio, C. \& Rijke, M. QRFA: A data-driven model of information-seeking dialogues. {\em Proc. ECIR}. pp. 541-557 (2019)
\bibitem{verberne2015user}Verberne, S., Sappelli, M., Järvelin, K. \& Kraaij, W. User simulations for interactive search: Evaluating personalized query suggestion. {\em Proc. ECIR}. pp. 678-690 (2015)
\bibitem{vlachou2022performance}Vlachou, M. \& Macdonald, C. Performance Predictors for Conversational Fashion Recommendation. {\em Proc. KaRS Workshop At RecSys}. (2022)
\bibitem{wang2023colbert}Wang, X., Macdonald, C., Tonellotto, N. \& Ounis, I. ColBERT-PRF: Semantic pseudo-relevance feedback for dense passage and document retrieval. {\em ACM Transactions On The Web}. \textbf{17}, 1-39 (2023)
\bibitem{wu2021fashion}Wu, H., Gao, Y., Guo, X., Al-Halah, Z., Rennie, S., Grauman, K. \& Feris, R. Fashion iq: A new dataset towards retrieving images by natural language feedback. {\em Proc. CVPR}. pp. 11307-11317 (2021)
\bibitem{wu2020fashion}Wu, H., Gao, Y., Guo, X., Al-Halah, Z., Rennie, S., Grauman, K. \& Feris, R. Fashion IQ: A New Dataset Towards Retrieving Images by Natural Language Feedback. {\em Proc. CVPR}. (2021)
\bibitem{wu2021partially}Wu, Y., Macdonald, C. \& Ounis, I. Partially Observable Reinforcement Learning for Dialog-based Interactive Recommendation. {\em Proc. RecSys}. (2021)
\bibitem{wu2022multi}Wu, Y., Macdonald, C. \& Ounis, I. Multi-Modal Dialog State Tracking for Interactive Fashion Recommendation. {\em Proc. RecSys}. (2022)
\bibitem{yu2017fine}Yu, A. \& Grauman, K. Fine-grained comparisons with attributes. {\em Visual Attributes}. pp. 119-154 (2017)
\bibitem{yu2019visual}Yu, T., Shen, Y. \& Jin, H. A visual dialog augmented interactive recommender system. {\em Proc. KDD}. (2019)
\bibitem{yu2020towards}Yu, T., Shen, Y. \& Jin, H. Towards hands-free visual dialog interactive recommendation. {\em Proc. AAAI}. (2020)
\bibitem{zhang2020evaluating}Zhang, S. \& Balog, K. Evaluating conversational recommender systems via user simulation. {\em Proc. KDD}. pp. 1512-1520 (2020)
\bibitem{zhang2022analyzing}Zhang, S., Wang, M. \& Balog, K. Analyzing and simulating user utterance reformulation in conversational recommender systems. {\em Proc. SIGIR}. (2022)
\bibitem{zhou2020towards}Zhou, K., Zhou, Y., Zhao, W., Wang, X. \& Wen, J. Towards topic-guided conversational recommender system. {\em ArXiv Preprint ArXiv:2010.04125}. (2020)
\bibitem{zou2019learning}Zou, J. \& Kanoulas, E. Learning to ask: Question-based sequential Bayesian product search. {\em Proc. CIKM}. (2019)

\end{thebibliography}






\end{document}